\documentclass[journal=jacsat,manuscript=article]{achemso}

\usepackage[version=3]{mhchem} 
\usepackage{float}
\usepackage{soul}
\usepackage{xcolor}

\author{Oren Elishav}
\altaffiliation{O.E. and O.B. contributed equally to this work}
\affiliation{Department of Chemical and Environmental Engineering, Yale University, New Haven, CT 06520, United States}
\alsoaffiliation{School of Chemistry, Tel Aviv University, Tel Aviv 6997801, Israel}
\author{Ofir Blumer}
\altaffiliation{O.E. and O.B. contributed equally to this work}
\affiliation
{School of Chemistry, Tel Aviv University, Tel Aviv 6997801, Israel}
\author{T. Kyle Vanderlick}
\affiliation{Department of Chemical and Environmental Engineering, Yale University, New Haven, CT 06520, United States}
\author{Barak Hirshberg}
\email{hirshb@tauex.tau.ac.il}
\affiliation
{School of Chemistry, Tel Aviv University, Tel Aviv 6997801, Israel}
\alsoaffiliation[Sackler]
{The Center for Computational Molecular and Materials Science, Tel Aviv University, Tel Aviv 6997801, Israel.}
\alsoaffiliation[Ratner]
{The Ratner Center for Single Molecule Science, Tel Aviv University, Tel Aviv 6997801, Israel.}
\alsoaffiliation{The Center for Physics and Chemistry of Living Systems, Tel Aviv University, Tel Aviv 6997801, Israel}

\title
  {The effect of ligands on the size distribution of copper nanoclusters: insights from molecular dynamics simulations}

\abbreviations{}

\begin{document}

\begin{abstract}
Controlling the size distribution in the nucleation of copper particles is crucial for achieving nanocrystals with desired physical and chemical properties. However, their synthesis involves a complex system of solvents, ligands, and copper precursors with intertwining effects on the size of the nanoclusters. We combine molecular dynamics simulations and DFT calculations to provide insight into the nucleation mechanism in the presence of a triphenylphosphite ligand. We identify the crucial role of the strength of the metal-phosphine bond in inhibiting the cluster's growth. We demonstrate computationally several practical routes to fine-tune the bond strength by modifying the side groups of the additive. Our work provides molecular insight into the complex nucleation process of 
 protected copper nanocrystals, which can assist in controlling their size distribution and, eventually, their morphology.  
\end{abstract}

\section{Introduction}
Metal nanoclusters (NCs) exhibit different physical and chemical properties as a function of their crystal morphology and size distribution.\cite{Liu2018} Therefore, control over the size and shape of the NCs during the nucleation is essential for their implementation in diverse applications in optoelectronics \cite{Lhuillier2015,Deng2012,Grim2014,Wang2012}, catalysis\cite{Sun2019,Somorjai2008,Elishav2022,FANG2016}, and sensing and imaging \cite{Ithurria2008,Joo2006,Chen2014,Gao2012,Homan2012,Li2013}. In recent decades, significant advancements have been achieved primarily in the realm of noble metals such as Au and Ag by reducing a metal precursor in the presence of organic ligands.\cite{Baghdasaryan2021,FANG2016,Cowan2020} More recently, researchers have explored effective synthetic protocols for alternative, earth-abundant metals such as copper.\cite{Kim2023,Reiss2016,DeTrizio2016,Lai2020} Therefore, understanding the nucleation and growth of cooper nanocrystals in different environments allows for finer control over their morphology. This allows a monodisperse starting material for nanoparticle production or to effectively develop ligand-protected copper nanoclusters.\cite{Kim2023,Lai2020} Also, insights into nanocluster aggregation are necessary for the conventional colloidal synthesis of inorganic nanoparticles.\cite{Kim2023,Stone2023}

Nanocluster synthesis usually involves adding ligands to precursors and solvents. Ligands can affect the size distribution and final shape of the NCs.\cite{Reiss2016} Therefore, new ligands are experimentally investigated by adding them directly during NCs' synthesis.\cite{Giansante2015} Ligands that form metal-phosphine complexes allow the bonding of the ligand group to the metal and thus alter the nucleation process in solution.\cite{Wei2022}  However, the landscape of possible ligands is too vast to be explored only by a trial-and-error experimental approach.\cite{Pascazio2022} 

MD simulations provide a microscopic view into the nucleation process as it evolves with time in atomistic detail~\cite{Sosso2016}, which can, in principle, aid the design of ligands that lead to a desired size distribution.\cite{Pascazio2022} Realistic MD simulations of crystallization and nucleation from solution are challenging due to the multi-component system's complexity and its long-time scales.\cite{Karmakar2019} Yet, in some cases, state-of-the-art MD simulations were performed and provided mechanistic insight or corroborated experimental results.\cite{Gebauer2014,Zahn2015,Harano2012,Davey2013,Dey2010,Zou2022,Finney2023} For example, MD simulations were used to understand the nucleation of metals, including Al, Fe, and Cu, from a melt.\cite{Mahata2018,Shibuta2015,Shibuta2017,Shibuta2010,Griffiths2021} Moreover, recent studies used MD  simulations to investigate ligand-metals interfaces and complexes.\cite{Pascazio2022,Zito2021,Li2017,Tiwari2023,Anushna2023}. Still, most computational studies are focused on noble metal nanoclusters, such as Au or Ag.\cite{Mirko2020,Tiwari2023,Dutta2021,BRANCOLINI2019,Yiyang2018}

In many experimental systems, ligands play a crucial role in controlling metal nanoclusters' stability and size.\cite{Mahsa2022,Evert2019} A recent experimental observation showed the significant effect of ligands on the resulting size distribution in the synthesis of copper nanoparticles.\cite{Stone2023} In that study, copper-ligand systems were investigated to control the crystallization and growth of copper nanoparticles. The synthesis procedure included mixing a precursor, solvent, and additive. One of the experimental procedures involves dissolving $0.5\,mmol$ of copper nitrate in $8\,ml$  Octadecene and $1\,gr$ Hexadecylamine at $250-260^\circ$C. Nucleation of copper in Octadecene showed large nanoparticles with an average diameter of $~200\,nm$. However, when $1.6\,ml$ of the ligand Triphenylphosphite (TPOP) was added to the solution, smaller nanoparticles were observed with an average diameter of $~10\,nm$.\cite{Stone2023}

In this paper, we use MD simulations and DFT calculations of this experimental model system to describe the ligand-protected copper nanocluster formation mechanism and the ligand influence on the resulting size distribution. We first provide the computational details and force field validation, followed by the nucleation simulation results.


\subsection{Computational details and model benchmarking}

We performed DFT calculations using the revPBE0 functional\cite{Zhang1998}, with the D3 dispersion correction and Becke-Johnson Damping \cite{Stefan2011}. We employed the 6-31G(d,p) basis \cite{Krishnan2008} and used Q-chem 6.0 \cite{qchem} to perform a geometry optimization and find the equilibrium bond length between a Copper (Cu) atom and a Phosphorus (P) atom of the TPOP molecule. We applied a basis set superposition error (BSSE) correction\cite{Balabin2008} on the optimized geometry of the complex to find the bond strength.

We performed MD simulations using the Large-scale Atomic/Molecular Massively Parallel Simulator, LAMMPS (30 Jul 2021)\cite{Thompson2022} with the Nos\'e-Hoover thermostat and barostat\cite{PhysRevB.69.134103} with a damping factor of 100 fs and 1000 fs for the temperature and pressure, respectively. The simulations were conducted with a timestep of $1\,fs$. Lennard-Jones (LJ) parameters for non-bonding interactions between different atoms or pseudo-atoms were calculated using traditional Lorentz-Berthelot combination rules\cite{10.1093/oso/9780198803195.001.0001}.
We employed the united-atom TraPPE force field (FF) for Octadecene~\cite{Wick2000} with four types of pseudo-atoms: $CH_3(sp^3)$, $CH_2(sp^2)$, $CH(sp^2)$ and an additional $CH_2(sp^2)$ type for the vinyl group. We tested this FF by calculating different properties in unbiased MD simulations at room temperature. We calculated the density in a simulation of a liquid with 1000 molecules and the bond lengths and angles in a simulation of a single molecule in a vacuum. The results showed a good comparison to experimental results (Table 1).

\begin{table} [H]
  \caption{Comparison between experimental and simulated values of density and bond length for Octadecene}
\label{tb2S:notes}
  \begin{tabular}{lll}
    \hline
    & Experiments & Simulation \\
    \hline
    Density $(g/mol)$  & 0.789 & 0.784±0.002\\
    Angles  &C-C=C 120\textsuperscript{\emph{a}} & 119.5±0.1\\ 
    &C-C-C  112.6±0.3\textsuperscript{\emph{a}}  & 113.8±0.1\\
    Bond $(pm)$  & C-C  154\textsuperscript{\emph{b}}&	154.05±0.02 \\ & C=C 133\textsuperscript{\emph{b}} & 133.00±0.08\\
    \hline
  \end{tabular}
  
  \textsuperscript{\emph{a}} experimental values obtained from ref~\cite{Chang1970}
  
  \textsuperscript{\emph{b}} experimental values obtained from ref~\cite{fox1995organic}
\end{table} 

 GAFF2 FF parameters for TPOP were calculated using Pysimm for an all-atom description. We then replaced all carbons and hydrogen atoms with CH pseudo-atoms to reduce the computational cost. For these pseudo-atoms, we chose LJ parameters for the phenol CH group in the TraPPE FF. We tested our model using unbiased simulations of TPOP, with 1000 molecules in a liquid when calculating the density and a single molecule in a vacuum when calculating optimal bond lengths (Table 2). 

\begin{table} [H]
  \caption{Comparison between experimental and simulated values of density and bond length for TPOP}
\label{tb2S:notes}
  \begin{tabular}{lll}
    \hline
    & Experiments & Simulation \\
    \hline
    Density $(g/mol)$  & 1.184 & 1.207±0.001\\
    Bond $(pm)$  & P-O 165\textsuperscript{\emph{a}} &	163.75±0.06\\
    &C-C 140\textsuperscript{\emph{b}} & 139.94±0.02\\ 
    & C-C 140\textsuperscript{\emph{b}} &140.09±0.03\\
    & O-C 136\textsuperscript{\emph{a}} & 137.60±0.07\\
    \hline
  \end{tabular}
  
  \textsuperscript{\emph{a}} experimental values obtained from ref~\cite{Golovanov2005}
  
  \textsuperscript{\emph{b}} experimental values obtained from ref~\cite{Hameka1987}
\end{table} 

We used the embedded atom model (EAM) FF to describe the Cu-Cu interactions. To test this model,
we evaluated the melting temperature of Cu using Metadynamics\cite{bussi_using_2020} simulations performed with PLUMED\cite{Tribello2014}. We used the ENVIRONMENTSIMILARITY\cite{Piaggi2019} collective variable with a bias pace of 500 timesteps, Gaussian height of $0.5\,k_BT$ and width of 3, and bias factor of 50. We retrieved the melting temperature of bulk Cu by performing simulations at different temperatures and calculating the free energy difference between the solid and liquid states (Figure 1a), showing that the EAM model provides a melting temperature of $~1325K$, close to the experimental value of $1357K$\cite{Cahill1962}. 
Interactions of Cu with other atoms were described with LJ interactions, found using Lorentz-Berthelot combination rules.
However, when applying these rules, LJ parameters for Cu-Cu interactions are required. We used parameters as suggested by Lv et al. \cite{Lv2011}. We emphasize that these LJ parameters were used only for calculating Cu-solution interaction. Nonetheless, we verified that these LJ parameters adequately describe Cu by calculating the radial distribution function of bulk Cu (Figure 1b). We compared it to the embedded atom model (EAM) and calculations by Neogi et al.\cite{Neogi2017} 
 
\begin{figure} [H]
  \includegraphics [scale=0.9] {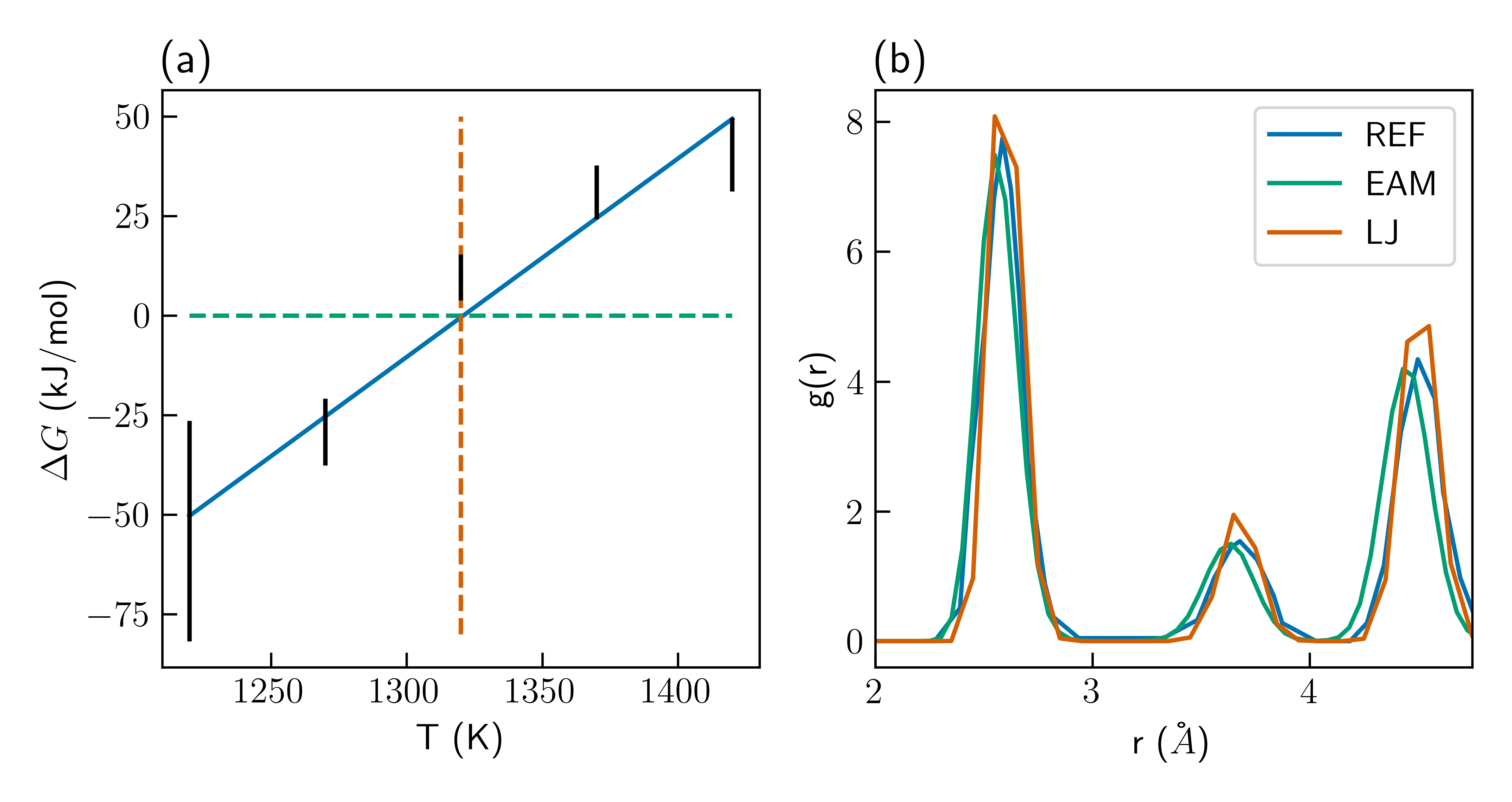}
  \caption{(a) Melting temperature calculation using EAM \cite{Foiles1986} for Cu-Cu interactions model. The orange line indicates the obtained melting temperature calculated at the intersection with $\Delta$G=0 (b) RDF calculation for Cu-Solution interactions using LJ \cite{Heinz2008} and comparison to simulation reference\cite{Neogi2017}.}
  \label{fgr:1}
\end{figure}

\section{Nucleation simulations}

To test our hypothesis that chemical bonds between TPOP and Cu are responsible for changing the size distribution of the nanoparticles, we performed MD simulations of nucleation at the relevant experimental temperatures. In our simulations, which used relatively high temperatures, nucleation happens spontaneously, and Metadynamics simulations are not required. 

The first simulation was performed in the isothermal-isobaric (NPT) ensemble at ambient conditions, including a premixed copper in solution without additive. The P-Cu bond strength was set to $15.3\, kcal \, mol^{-1}$, based on preliminary DFT calculation of the complex (see more details below). A snapshot from the simulations is shown in (Figure 2a), which exhibits nucleation to a single nanocluster. However, in the presence of the additive TPOP, significant multiple smaller nanoclusters are observed, as shown in (Figure 2b). Each copper nanocluster is surrounded by several TPOP molecules, as presented in (Figure 2c). The P atoms bond to the copper atoms, while the residues of the TPOP are oriented toward the outer side of the nanocluster. Thus, the TPOP molecules inhibit the growth of the copper nanocluster and allow a smaller copper cluster size distribution. The findings were corroborated by three independent trajectories. 

\begin{figure} [H]
  \includegraphics [scale=0.48] {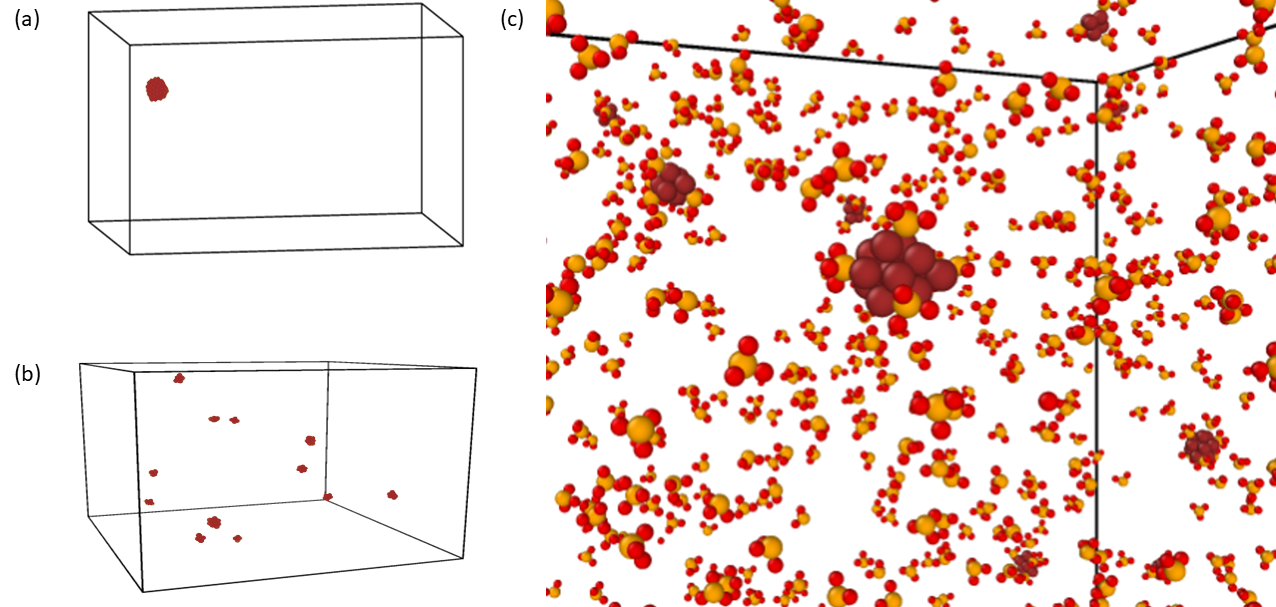}
  \caption{MD simulation of copper nucleation: (a) Copper nanoclusters without TPOP (for clarity, only copper atoms are shown), (b) Copper nanoclusters with TPOP (for clarity, only copper atoms are shown), and (c) snapshot of copper nanocluster and TPOP molecules (for clarity only Cu, P, and O atoms are shown in burgundy, orange, and red).}
  \label{fgr:2}
\end{figure}

We tested the effect of the bond strength between the P and Cu atoms on the number of copper nanoclusters formed and their size, as shown in Figure 3. As the P-Cu bond strength decreases, the average number of atoms in the Cu nanoclusters increases, and the overall number of copper clusters decreases. Thus, altering the P-Cu bond strength can provide fine-tuning for the size distribution of the produced copper clusters. For example, modifying the ligand by adding an electrons withdrawing group can weaken the P-Cu bond. The weaker P-Cu bond allows copper clusters to grow and increase in size by merging smaller clusters. In addition, we tested the effect of temperature on the number of nanoclusters during the nucleation simulation. We performed simulations at temperatures of $190,210,230$ and $250^\circ$C. As the temperature decreases, the number of nanoclusters after $400ns$ slightly increases (Figure S1). Lower temperature decreases the diffusion of Cu atoms, thus hindering the formation of larger clusters, but the effect is less profound than the P-Cu bond strength.    

\begin{figure} [H]
  \includegraphics [scale=0.9] {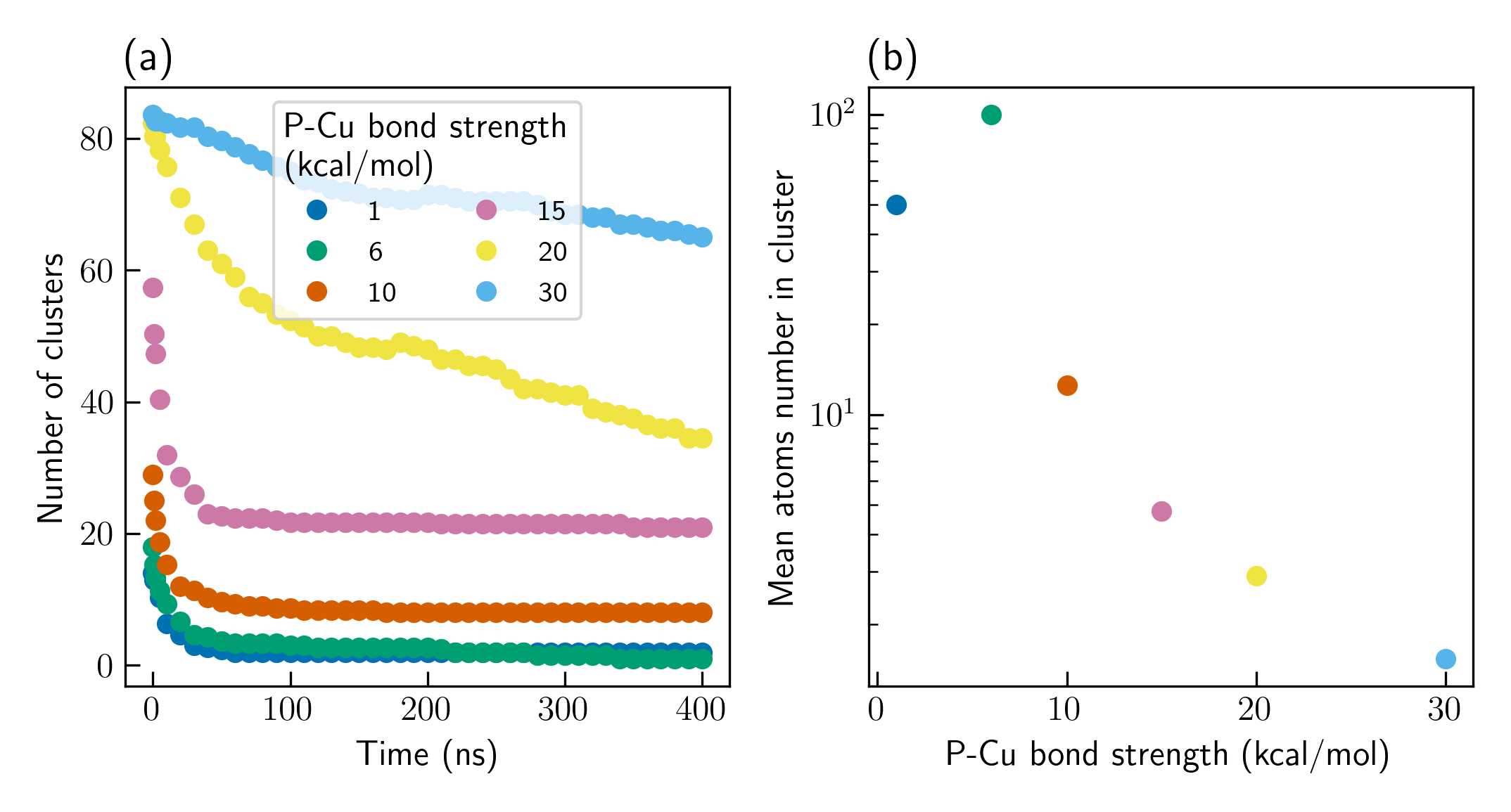}
  \caption{(a) Number of observed copper nanoclusters as a function of time for various P-Cu bond strengths, and (b) Mean number of atoms in a nanocluster after $400\,ns$ of MD simulation, as a function of bond strength. Colors correspond to the same bond strength as in the legend of panel (a).}
  \label{fgr:3}
\end{figure}

Using DFT calculations, we tested some practical approaches to increase the P-Cu bond strength experimentally. We looked at the effect of swapping two hydrogen atoms with electron-donating (-NH2) groups (Figure 4b). We also tested how swapping the phenyl residue (Figure 4c) with a smaller (methyl) residue affects the bond strength. We optimized the structure of the complexes with the suggested ligands and performed harmonic vibrational frequency calculations to confirm that they are local minima with no imaginary frequencies. Figure 4 shows the highest occupied molecular orbitals (HOMO) and the lowest unoccupied molecular orbitals (LUMO) for the original and altered complexes. 

Swapping only hydrogen atoms with an NH2 group (Figure S2a-c) only slightly increases the bond strength ($15.6 kcal \, mol^{-1}$). Yet, when swapping two hydrogen atoms on different phenyl residues, as shown in Figure 5b, the bond strength increases to $19.8 kcal \, mol^{-1}$. Swapping three hydrogen atoms with NH2 groups on different phenyl rings (Figure S2d-f) achieves a similar bond strength ($20.5 kcal \, mol^{-1}$).  In the latter case, the molecular conformation of the complex is also very different relative to the original ligand (Figure 4a,b). When replacing one phenyl ring with a methyl group, the bond between the copper and the altered TPOP increases even further to $28.8 kcal \, mol^{-1}$,and the conformation of the complex changes (Figure 4a,c). Changing the phenyl ring with a smaller residue will also affect the evaporation of the ligand during the high-temperature synthesis process. Thus, only one phenyl ring was replaced to avoid significant changes in the physical properties. The suggested examples of modifying the ligand provide practical ways of controlling the complex binding energy and, as a result, the nanoparticle size distribution through additive design guided by MD simulations.

\begin{figure} [H]
  \includegraphics [scale=0.6] {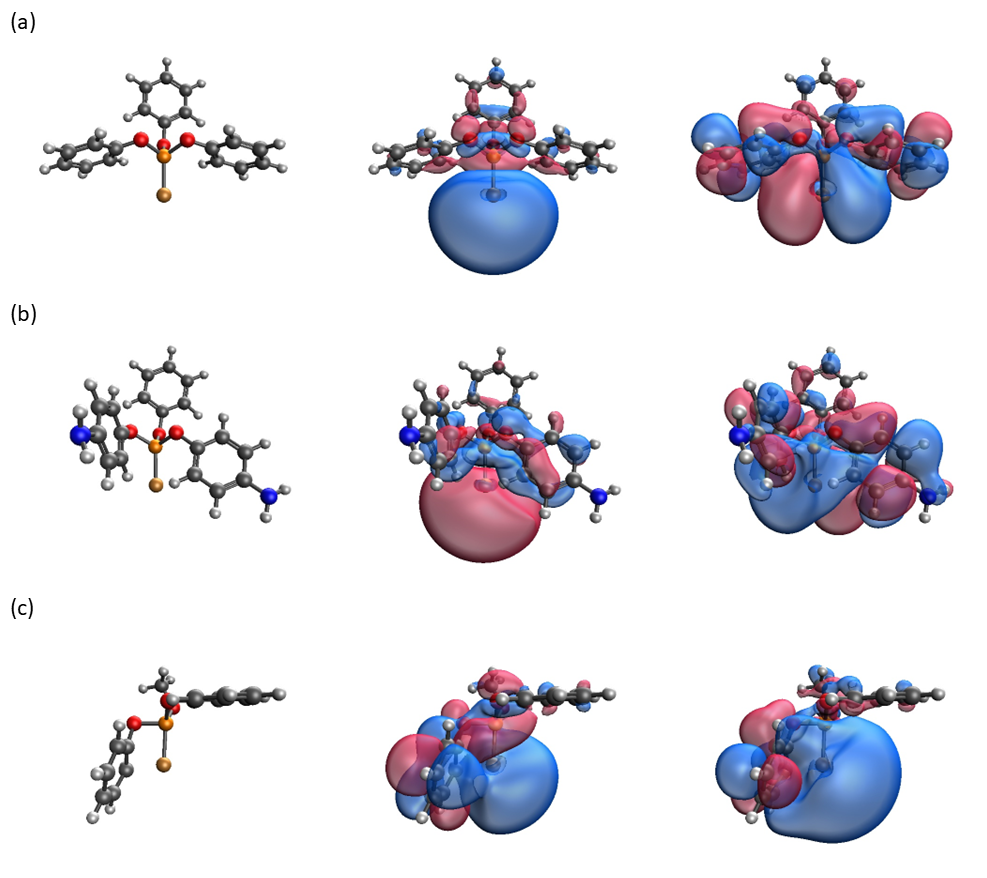}
  \caption{DFT calculation of TPOP additive with Cu atom, optimized geometry, HOMO and LUMO (a) original molecule, (b) modified TPOP with electrons donating group (-NH2), and (c) modified TPOP with methyl group. Atom colors of H, C, O, P, Cu, and N are white, gray, red, orange, gold, and blue, respectively.}
  \label{fgr:4}
\end{figure}

\section{Conclusions}
To summarize, we investigated the nucleation of ligand-protected copper nanoclusters in order to design ways to control their size distribution. We showed that adding a TPOP additive resulted in smaller NCs relative to a system without it. We used MD simulation to follow the nucleation process and understand the mechanism leading to control over NCs size. The metal-phosphine bond of copper atoms and ligand blocks the coalescence and growth of copper clusters, thus enabling smaller NCs. We showed that decreasing the P-Cu bond strength resulted in larger copper nanoclusters. P-Cu bond strength equal to or below $6 kcal \, mol^{-1}$ resulted in a single nanocluster, similar to the nucleation process without the additive. We used DFT calculations to examine the effects of modifying the side group of TPOP on the P-Cu bond strength. Altering the ligand with such groups can allow fine-tuning of the copper nanocluster size, which is interesting to test experimentally. Our results show the potential of MD simulations in aiding the design of improved ligands for controlling the size distribution of nanoclusters with desired properties.

\begin{acknowledgement}
B.H. Acknowledges support by the Pazy Foundation (Grant No. 415/2023), the USA-Israel Binational Science Foundation (Grant No. 2020083), and the Israel Science Foundation (Grants No. 1037/22 and 1312/22). O.E. thanks the Israel Academy of Sciences and Humanities Postdoctoral Fellowship Program for Israeli Researchers, the Faculty of Exact Sciences Dean’s Fellowship, and the Ratner Center Fellowship at Tel Aviv University.

\end{acknowledgement}

\begin{suppinfo}
Supporting Information: Additional computational details, including LAMMPS computational details, the effect of temperature on nanocluster size, other TPOP molecules modified with NH2 groups, and the GitHub repository for simulation input files.

\end{suppinfo}

\bibliography{refs}

\end{document}


\section{Computational details}

All molecular dynamics (MD) simulations were performed in the Large-scale Atomic/Molecular Massively Parallel Simulator\cite{Thompson2022} (LAMMPS), with a timestep of 1 fs. The creation of the initial state of the system followed the protocol described below. First, a “primary” cell was created with a ratio of 1:8:40 (Cu:TPOP: Octadecene) or
1:48 (Cu:Octadecene). It was replicated in the x,y,z coordinates by a factor 5X5X4 to gain a system with 100 Cu atoms.
Then, we initialized a mixing stage of 2 ns at 300K (NPT, 1 bar), followed by 3 ns of heating (NPT, 1 bar). Finally, we executed a
thermalisation of 1 ns at 520K (NPT, 1 bar).
Production simulations also maintained a temperature of 520K and 1 bar (NPT) pressure. Input files for MD simulation are available at https://github.com/oreneli/Copper-nucleation.git

\begin{figure} [H]
  \includegraphics [scale=0.45] {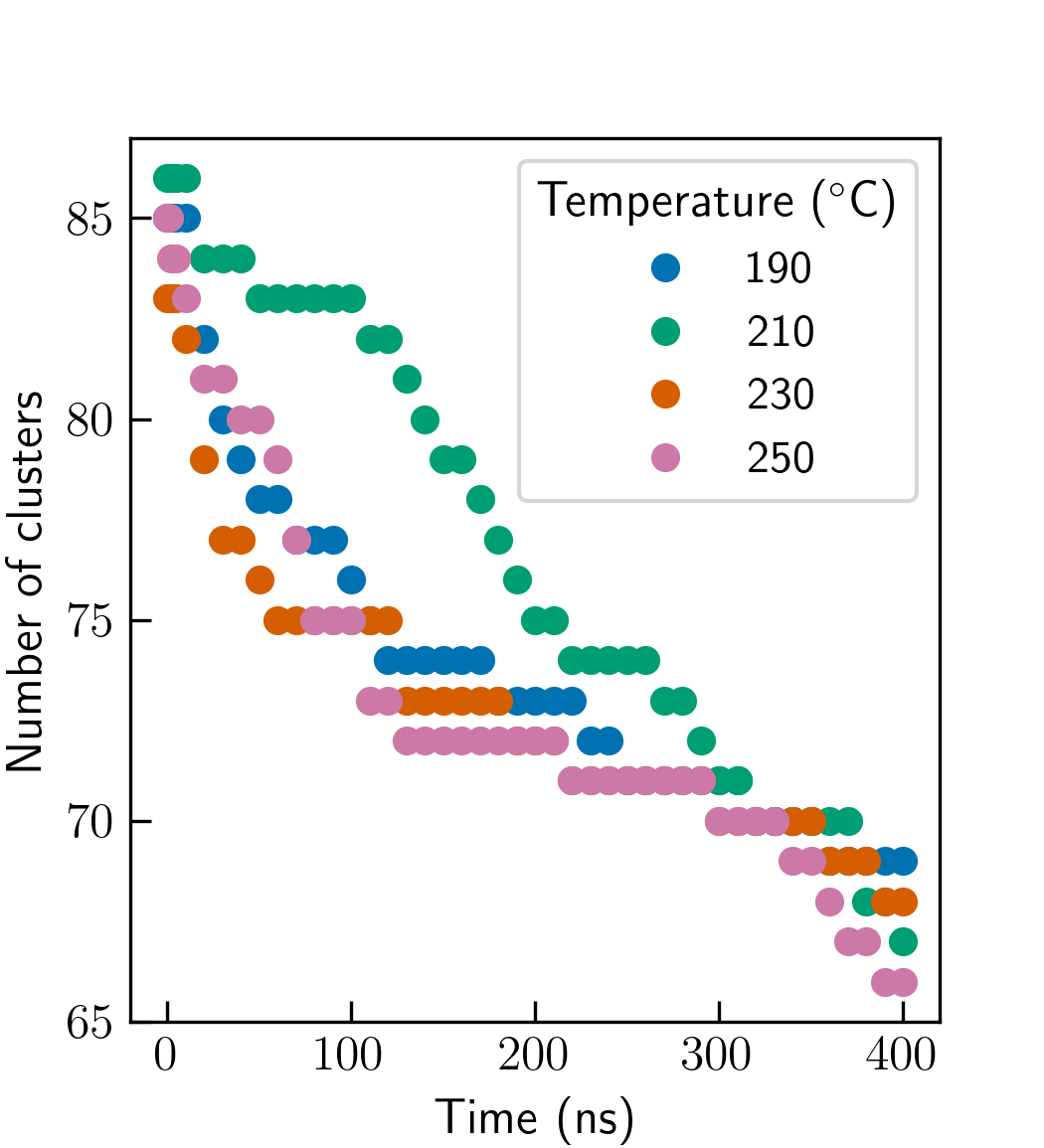}
  \caption{Number of observed copper clusters as a function of temperatures}
  \label{fgr:1}
\end{figure}

\begin{figure} [H]
  \includegraphics [scale=0.6] {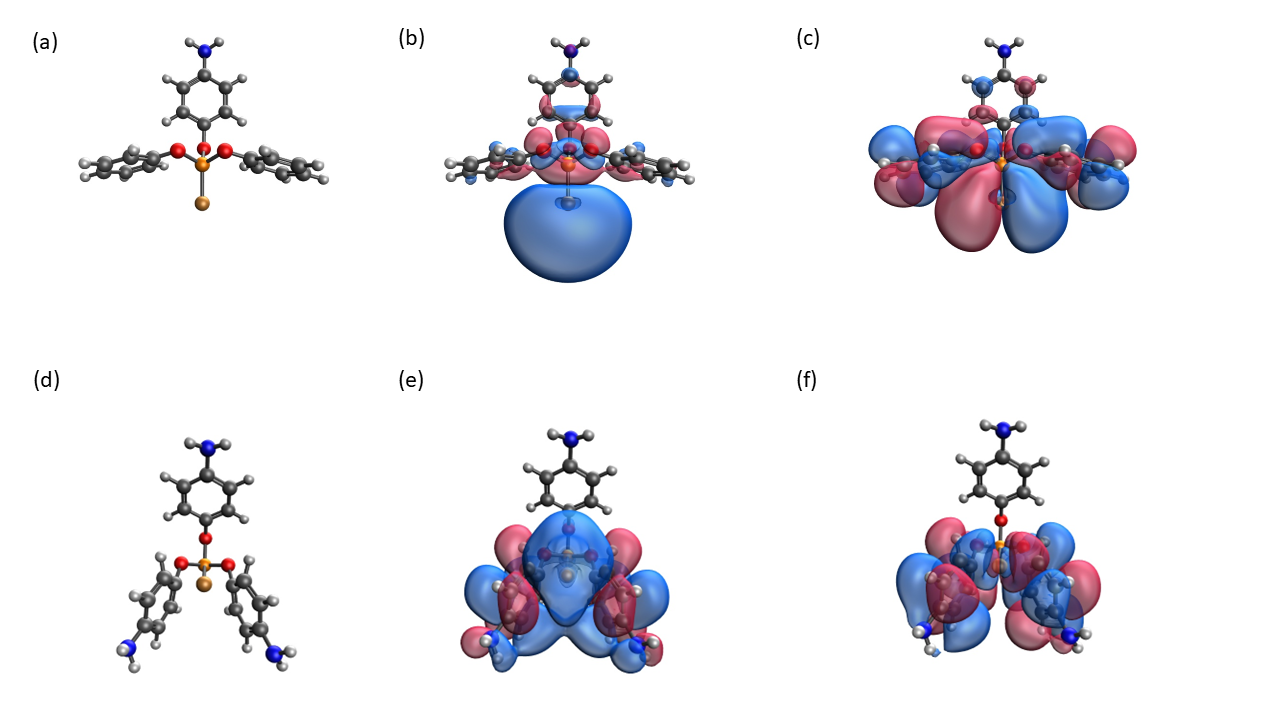}
  \caption{DFT calculation of modified TPOP with electrons donating group (-NH2) with Cu atom with a single NH2 group swap (a) optimized geometry, (b) HOMO, (c) LOMO, with three NH2 groups swap (d) optimized geometry (e) HOMO, and (f) LOMO}
  \label{fgr:2}
\end{figure}

\bibliography{refsSI}